# A dynamical model of a GRID market

Uli Harder, Peter Harrison, Maya Paczuski & Tejas Shah
Department of Computing & Department of Mathematics
Imperial College London, South Kensington Campus
London SW7 2AZ, United Kingdom
{uh, pgh}@doc.ic.ac.uk, maya@imperial.ac.uk, tss101@doc.ic.ac.uk

## Abstract

*We discuss potential market mechanisms for the GRID. A complete dynamical model of a GRID market is defined with three types of agents. Providers, middlemen and users exchange universal GRID computing units (GCUs) at varying prices. Providers and middlemen have strategies aimed at maximizing profit while users are 'satisficing' agents, and only change their behavior if the service they receive is sufficiently poor or overpriced. Preliminary results from a multi-agent numerical simulation of the market model shows that the distribution of price changes has a power law tail.*

## 1 Introduction

From the hardware point of view the GRID is a collection of computing, storage and networking devices. At first glance it looks like standard cluster computing. However, the GRID is equipped with middleware that makes it easy for the user to find resources. A good analogy is the Internet where search engines are the "information middleware" that bring clients and services together [11].

There are currently two main options for middleware: Globus [1] and Legion [2]. Ideally the GRID should be a heterogeneous collection of computing equipment. At present it runs on LINUX machines.

Currently only a few GRIDs exists. In the UK there are two main GRIDs, one is used by computer scientists to develop middleware and architectures. The other, GridPP, is used particle theorists. Soon this effort will be enlarged to cope with the data created by the Large Hadron Collider at CERN. Other scientists are starting to use GRID technology for their purposes. Another pool of potential users are banking and insurance companies where computational applications are intensive. Further afield but no less important are game companies beginning to look at how they can use GRID technology.

## 2 The future of the GRID

A consensus exists in the research community that the GRID, if it is successful, will be used in a commercial context. For instance, companies may specialise in providing GRID services in a similar way telephone companies operate today. A second scenario is that large organizations will sell spare computing power to others. A third scenario sees private PC owner selling spare CPU cycles of their machines. Since the GRID is, in a way, meant to do away with the PC this seems a less likely option. In this paper we present a market model for the first scenario.

## 3 Charging for computing time

The name "GRID" was chosen in analogy with the electrical power grid. Just as electrical appliances are plugged into a socket, computing applications are plugged into the GRID. So, it is only a matter of time until GRID providers will want to sell computing power to end users. In that case, providers will develop charging schemes for their services in a competitive environment.

We expect that, similar to electricity markets, few negotiations will occur between producers and end users. In the privatised electricity market, middlemen buy electricity from producers and sell it to end users through a range of service plans and tariffs. In fact, detailed multi-agent simulations of the UK and other electricity markets have been made [4].

For the GRID, an additional complication arises since consumption can not be measured easily. The user needs several components to perform a computation: CPU, disks, memory and networking. If one of these components is missing all others loose their value. For simplicity, we assume that in the future a universal "GRID computing unit" (GCU) will be established. In fact this may simply be measured in real time access like telephone usage. In this case middlemen would make contracts that specify the minimum number of CPUs, memory, etc. that they supply for specific time periods. Payments in the GRID-world could be made in a similar way utility/phone bills are settled nowadays.

Charging for computer access, or indeed any commodity, service, etc., can have two different goals:

- maximisation of utilization, or fair access to machines that have been purchased collectively

- profit maximisation.

In this paper we assume that providers and middleman endeavor to maximise their profit using some strategies, or rules of thumb. End users, on the other hand, are not viewed as utility maximizing agents, but rather as 'satisficing' agents who maintain their behavior unless they become dissatisfied, a description of human economic behavior put forward Simon [19]. These agents want to lessen their effort and do not make detailed investigations into the market. Producers, middlemen and users interact via auctions or commodity markets, or a combination of both.

## 4 Background

There is a substantial body of literature in the statistical physics community for agent based models of markets. For example, Bak, Paczuski and Shubik [6] made simple models of the stock market and found that large variations in prices were due to a crowd effect, where agents imitate each other's behavior. More elaborate models of the London stock exchange have been made by Farmer and collaborators [24, 13]. Bak, Norrelykke and Shubik [5] found that the value of money, which is not determined (or even addressed) in equilibrium economic theory, is due to a dynamical symmetry breaking in networks of interacting agents with bounded rationality. Challet and Zhang [10] introduced the 'minority model' as a perspicuous adaptation of Arthur's "El Farol Bar Problem" [3], where agents with bounded rationality compete for limited resources and their strategies co-evolve. Nagel, Shubik and Strauss [14] showed that a separation of time scales if often needed for a well-defined market to self-organize. It is with this general background that we approach the problem of modelling a potential GRID market.

The oldest record of a charging system for computer access that we are aware of is from Sutherland in 1968 [21]. A large body of literature from the 70's looked into charging for access to mainframe computers [15, 12].

More recently authors have begun to investigate a possible commodity market created by the GRID. There is "G-commerce" [23] which looks at auctions and commodity markets. The "Compute Power Market" approach is more focussed on a completely heterogeneous market with supplier and consumers of all sizes [8]. There is also cpu charging (SPAWN) [22] and Popcorn [17] and a scheme based on CUMULUS [20]. Authors have also explored auctioning systems as replacements for scheduling algorithms [9]. For completeness we mention ideas from the mobile agent community where financial ideas are used to manage mobile agents [7].

## 5 Our model

In our model there are three types of agents: the GRID, middlemen and users. Though the GRID is likely to be at least several organizations, in our present model there is only one GRID provider. The GRID provides GCU's to the collection of middlemen at a constant rate. Each middleman passes his GCUs to his user base. Middlemen do not exchange with other middlemen and users cannot interact with the GRID directly. This setup is very similar to the Resource Allocation Game [18], an extension of the minority model [10]. However, in this case we have the additional structure of a market and a charging scheme. Our model has two time scales, a fast one and a slow one.

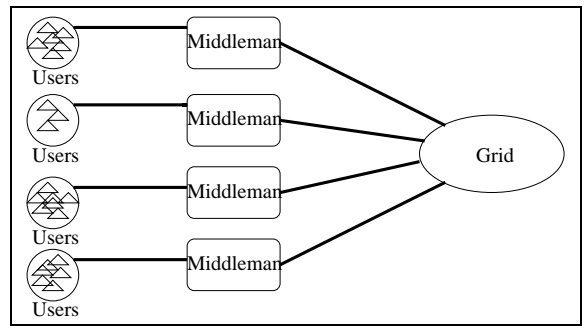

**Figure 1. Diagram of the model.**

With each *tick* at the fast time scale, the GRID produces a fixed number of GCUs. These are auctioned off to the middlemen who put in a bid for the number of GCUs they expect to need at the next tick to satisfy the demand of their users. The GRID allocates the GCUs to the middlemen by serving the highest bids first. If there is a greater demand than supply, some middlemen do not get any or all of the GCUs they bid for, and do not pay for unprovided GCUs. If there is oversupply, the GRID destroys GCUs. The number of GCUs each middleman bids for is determined by the past demand of his current users. He bids all his cash supply except a small safety margin. So the price each middleman is prepared to pay for a single GCU is simply his cash supply, minus a safety margin, divided by the number GCUs he is bidding for.

Each user may demand one GCU from her respective middleman at each tick. Users are characterized by the probability $q$ that they demand use, which is chosen from a uniform distribution in the unit interval. If a user requests a GCU, she pays her middleman immediately at the agreed price if she actually receives the GCU. If the middleman cannot deliver a GCU, because he has got none left, there is no charge for the demand. The middlemen use the money they collect this way for their next bid to the GRID.

At the slow time scale, each middleman decides whether to change his current price $p_t$ for providing GCUs to his users. His decision depends on the cash supply he currently holds $C_t$, the amount of cash he started out with initially, $C_0$, and the current average price $\langle p \rangle_t$ offered by the middlemen:

$$p_{t+1} = \begin{cases} p_t/2 & \text{if } C_t > 2C_0 \\ p_t & \text{if } C_0 < C_t \leq 2C_0 \\ \langle p \rangle_t & \text{if } C_t < C_0. \end{cases}$$

The amount of initial cash $C_0$ can vary from middleman to middleman, but must be positive.

The first rule stops inflation and might be thought of as intervention by a regulator or the tax authorities. The second rule preserves the status quo (*don't fix it if it ain't broken*). The last rule is likely to make the middleman's price more attractive to users.

At the same time users decide whether to change middlemen. Their decision is based on two indicators: service quality and price change. At each tick they make a note whether their demand was satisfied and compute their service quality, $S$, as the fraction of ticks their demand was satisfied. (Middlemen may run out of GCUs if they have not secured enough in the GRID auction.) The average service level achieved by all brokers is $\langle S \rangle$. A user changes her middleman if

$$sTol < \frac{\langle S \rangle - S}{\langle S \rangle}$$

where $sTol$ is the tolerance level of a particular user. Similarly, users dislike price hikes. If $P^{new}$ is the new price offered by her middleman, $P^{old}$ the previous price and $\langle P \rangle$ the previous average price over all middlemen, a user will change her middleman if

$$cTol < \frac{P^{new} - P^{old}}{\sqrt{\langle P \rangle P^{old}}}$$

where $cTol$ is her price tolerance level. Each user has her own price and service tolerance level which are fixed throughout the simulation. If a user is unsatisfied according to either criteria, she picks a new middleman at random.

In numerical simulations, we used 10,000 ticks at the fast time scale for every update at the slow time scale. This roughly approximates the scale separation between minutes and months.

## 6 Preliminary Results

We implemented the model described above as a multi-agent simulation. To get some indication how the model behaves we ran a configuration with 10 middlemen and 100 users demanding 50 GCUs on average whilst the GRID produces 40, 45 and 50 GCUs. The tolerance levels for price changes and service were set to 0.5 for all users. The initial cash supply for all middlemen was set to $C_0 = 21$. We consider the average prices offered by the middlemen and look at the distribution of changes in this market index from one update at the long time scale to the next. As shown in Figures 2 and 3, the distribution is very broad indicating a high volatility of the market. In fact a double-logarithmic plot reveals that the price changes are scale invariant with a critical index $\approx 1.3$, as shown in Figure 3.

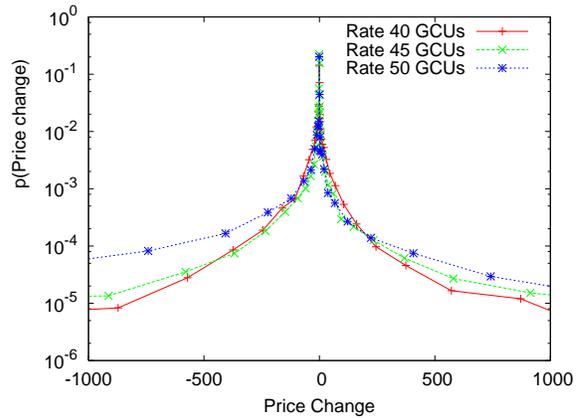

**Figure 2. A half-logarithmic plot of the probability distribution of market average price changes.**

When an oversupply of GCUs exists, the model quickly reaches a monopoly state with all users trapped by one middleman, as their service level is satisfactory and the price does not change any further. In the case of oversupply the price becomes fixed in time, while for undersupply the price exhibits a complicated dynamics. The balance between supply and demand rates can therefore be considered as a type of absorbing state phase transition point [16]. Ideally, one would like to make supply and demand balanced by market forces, or "self-organized" in the model rather than preset.

## 7 Future work

Future investigations into models for the GRID market need to consider simple models to understand what re-ally drives the system, as well as look at different auction schemes and ways to model user behavior. Additionally we need a more satisfactory pricing mechanism, perhaps like ones used in the electricity markets.

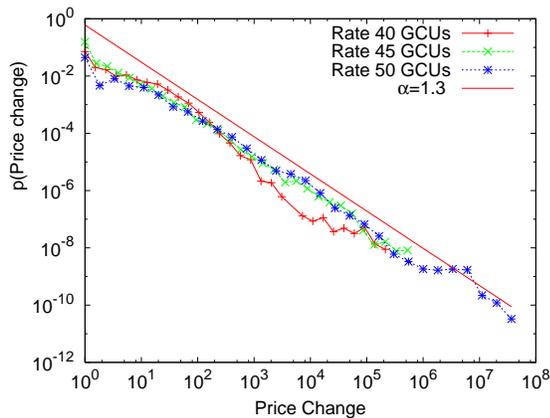

**Figure 3. A double logarithmic plot of the distribution of positive price changes. The price changes exhibit a fat tail with a power law index, or slope on the log-log scale $\approx -1.3$.**

## Acknowledgments

The research was partially funded by EPSRC (research grant PASTRAMI, GR/S24961/01).

## References

[1] Globus. www.globus.org.

[2] Legion. *www.cs.virginia.edu/˜legion/*.

[3] W. B. Arthur. Inductive Reasoning and Bounded Rationality (The El Farol Problem). *American Economic Review (Papers and Proceedings)*, 84:406–411, 1994.

[4] A. Bagnall. *Applications of Learning Classifier Systems, Studies in Fuzziness and Soft Computing vol. 150*, chapter A Multi-Agent Model of the UK Market in Electricity Generation. 2004.

[5] P. Bak, S. Norrelykke, and M. Shubik. Dynamics of money. *Physical Review E*, 60(3):2528–2532, 1989.

[6] P. Bak, M. Paczuski, and M. Shubik. Price Variations in a Stock Market with Many Agents. *Physica A*, 246:430, 1997.

[7] J. Bredin, R. T. Maheswaran, Ç. Imer, T. Başar, D. Kotz, and D. Rus. A game-theoretic formulation of multi-agent resource allocation. In *Proceedings of the Fourth International Conference on Autonomous Agents*, pages 349–356. ACM Press, June 2000.

[8] R. Buyya. Economic-based Distributed Resource Management and Scheduling for Grid Computing. *PhD Thesis, Monash University, Melbourne, Australia, April 2002*. cs.DC/0204048.

[9] R. A. Cagliano, M. D. Fraser, and M. E. Schaefer. Auction Allocation for Comuting Resources. *Communications of the ACM*, 38(6):88–96, 1995.

[10] D. Challet and Y.-C. Zhang. Emergence of Cooperation and Organization in an Evolutionary Game. *Physica A*, 246:407, 1997.

[11] P. Clarke. Introduction. *Getting ready for the Grid 11 February 2004 Institute of Physics HEPP group*, 2004. http://www.gridpp.ac.uk/meetings/iopfeb04/.

[12] I. W. Cotton. Microeconomics and the Market for Computer Services. *Computing Surveys*, 7(2):95–111, 1975.

[13] J. D. Farmer, P. Patelli, and I. I. Zovko. The Predictive Power of Zero Intelligence in Financial Markets. Technical report, http://lanl.arxiv.org/abs/cond-mat/0309233, 2003.

[14] K. Nagel, M. Shubik, and M. Strauss. The importance of timescales: simple models for economic markets. *Physica A*, 340(4):668–677, 15 September 2004.

[15] N. R. Nielsen. The Allocation of Computing Resources – Is Pricing the Answer? *Communications of the ACM*, 13(8):467–474, 1970.

[16] M. Paczuski, S. Maslov, and P. Bak. Field-theory for a model of self-organized criticality. *Europhys. Lett.*, 27:97, 1994.

[17] O. Regev and N. Nisan. The POPCORN Market an Online Market for Computational Resources, 98. *Proceedings of the first international conference on Information and computation economies, pp. 148 - 157 Series-Proceeding-Article Year of Publication: 1998 ISBN:1-58113-076-7*.

[18] R. Savit, S. A. Brueckner, H. V. D. Parunak, and J. Sauter. Phase structure of resource allocation games. *Physics Letters A*, 311(4-5):359–364, 19 May 2003.

[19] H. Simon. *Models of Man*, chapter A Behavioral Model of Rational Choice. 1957.

[20] B. Stiller, J. Gerke, P. Flury, P. Reichl, and Hasan. Charging Distributed Services of a Computational Grid Architecture. *Computer Engineering and Networks Laboratory TIK, Swiss Federal Institute of Technology, ETH Zurich, Switzerland*.

[21] I. E. Sutherland. A futures market in computer time. *Commun. ACM*, 11(6):449–451, 1968.

[22] C. A. Waldspurger, T. Hogg, B. A. Huberman, J. O. Kephart, and W. S. Stornetta. Spawn: A distributed computational economy. *Software Engineering*, 18(2):103–117, 1992.

[23] R. Wolski, J. S. Plank, J. Brevik, and T. Bryan. Analyzing Market-based Resource Allocation Strategies for the Computational Grid. *The International Journal of High Performance Computing Applications, Sage Science Press*, 15(3):258–281, 2001.

[24] I. Zovko and J. Farmer. The power of patience: A behavioral regularity in limit order placement. *Quantitative Finance*, 2(5):387–392, 2002.